\documentclass[aps,prx,twocolumn,showpacs,superscriptaddress,longbibliography]{revtex4-1}

\usepackage{amssymb}
\usepackage{amsfonts}
\usepackage{amsmath}
\usepackage{bm}
\usepackage{textcomp}
\usepackage{color}
\usepackage{longtable}
\usepackage{graphicx}
\usepackage{dcolumn}
\usepackage[a4paper=true,pagebackref=false]{hyperref}
\usepackage[normalem]{ulem}
\usepackage{physics}
\usepackage[T1]{fontenc}

\usepackage{xr}

\makeatletter

\newcommand*{\addFileDependency}[1]{
\typeout{(#1)}
\@addtofilelist{#1}
\IfFileExists{#1}{}{\typeout{No file #1.}}
}\makeatother

\newcommand*{\myexternaldocument}[1]{%
\externaldocument{#1}%
\addFileDependency{#1.tex}%
\addFileDependency{#1.aux}%
}

\myexternaldocument{./supplement/main}

\begin{document}


\title{
Intrinsic Quasiparticle Lifetime in a Superconducting Aluminum
}

\author{K.~Norowski}
\email{norowski@ifpan.edu.pl}
\affiliation{CoolPhon Group, International Research Centre MagTop, Institute of Physics, Polish Academy of Sciences, Aleja Lotnikow 32/46, PL-02668, Warsaw, Poland}

\author{M.~Foltyn}
\affiliation{CoolPhon Group, International Research Centre MagTop, Institute of Physics, Polish Academy of Sciences, Aleja Lotnikow 32/46, PL-02668, Warsaw, Poland}

\author{A.~Savin}
\affiliation{QTF Centre of Excellence, Department of Applied Physics, Aalto University, FI-00076, Aalto, Finland}

\author{M.~Zgirski}
\email{zgirski@ifpan.edu.pl}
\affiliation{CoolPhon Group, International Research Centre MagTop, Institute of Physics, Polish Academy of Sciences, Aleja Lotnikow 32/46, PL-02668, Warsaw, Poland}

\date{\today}

\begin{abstract}{
We use time-resolved thermometry to monitor the decay of nonequilibrium quasiparticles in superconducting Al in the temperature range from $0.3~$K to $1.2~$K. The quasiparticle lifetime at higher temperatures ($T>0.7~$K) agrees well with the calculated energy flow from electrons to phonons, but at lower temperatures it is significantly shorter than the theory predicts. We show well-defined internal equilibrium of quasiparticle system in the studied thermal transients, which implicates that quasiparticle-quasiparticle relaxation is much faster than electron-phonon interaction.}
\end{abstract}






\maketitle

Quasiparticles (QPs) are the most fundamental excitations present in the superconductor at a finite temperature. The excess QPs arise from absorption of photons travelling in wires probing the sample \cite{Cleland2004,Pekola2009ElectronicRefrigeration}, phonons propagating in the insulating substrate \cite{Nichele2022,Gasparinetti2024}, diffusion from regions of the sample where the dissipative elements have been probed with electric current \cite{Zgirski2020,Nahum1998,Reulet2016}, relaxation of two level fluctuators \cite{Delsing2022}, moving superconducting vortices \cite{Zgirski2024,Greenblatt2001}, phase slips \cite{Winkelmann2023}, or absorption of cosmic rays \cite{Oliver2020}. Very often they are perceived as having detrimental undesired effect on quantum circuits such as qubits \cite{Glazman2012}, resonators \cite{Klapwijk2014}, single electron transistors \cite{Pekola2012APL} and microcoolers \cite{Pannetier2009}. On the other hand they make possible operation of superconducting bolometers, e.g. single photon nanowire detectors \cite{Zwiller2021} or transition edge sensors \cite{Joel2013}. QPs may also be injected into a nanowire from a voltage-biased gate providing a mean for controlling critical current of such devices \cite{DeSimoni2018, Pashkin2021,Chapelier2023,Gasparinetti2024}. On top of that, quantum properties of superconductors can be in some cases mimicked by the effects of the thermal origin, governed by the quasiparticle population \cite{Berggren2018,Schwall2005}. It is thus important to study lifetimes of QPs in superconductor, $\tau_{qp}$, particularly in aluminum being a material of choice for many quantum devices. The literature provides direct measurements of $\tau_{qp}$ in aluminum only well below $T_c$ \cite{Wilson2001,Klapwijk2008,Klapwijk2011,deVisser2021, Klapwijk2009}. Owing to the superb temporal resolution of the switching thermometry \cite{Zgirski2018} we are able to measure $\tau_{qp}$ almost all the way up to the critical temperature $T_c$. We compare our data against the numerical model assuming two relaxation channels for the excess electron energy: the electron-phonon coupling and the QP diffusion. In addition we confront our data with QPs relaxation times recorded for $T<0.3~$K by other research groups.

The set of experiments made on diffusive conductors in the end of previous century showed that QPs in a normal state strongly interact with each other creating quasi-equilibrium described by Fermi-Dirac distribution \cite{Devoret1996,Rooks1997}. Particularly, it was shown that in a metal the electron-electron interaction time $\tau_{ee}$ is much shorter than electron-phonon relaxation time, $\tau_{ep}$ \cite{Altshuler1982}. The deviation from the Fermi-Dirac were observed \cite{Devoret1997,Esteve2003}, but still consistent with conclusion that $\tau_{ee} \ll \tau_{ep}$. The number of QPs goes down dramatically with temperature. It has two consequences. First, $\tau_{ee}$ in superconductor becomes larger than in the normal state, for it is more difficult for a non-equilibrium QP to find a mate to "agree" on the Fermi-Dirac distribution. Secondly, the ability of QPs to loose energy as a whole to phonon subsystem in the process of recombination back into Cooper pairs becomes suppressed as the temperature goes down towards absolute zero. Which process dominates? Is $\tau_{ee}$ still much shorter than $\tau_{ep}$ in a superconducting state? It is the question which we experimentally address in the current paper.

\begin{figure*}[t]
\centering
\includegraphics[width=\textwidth]{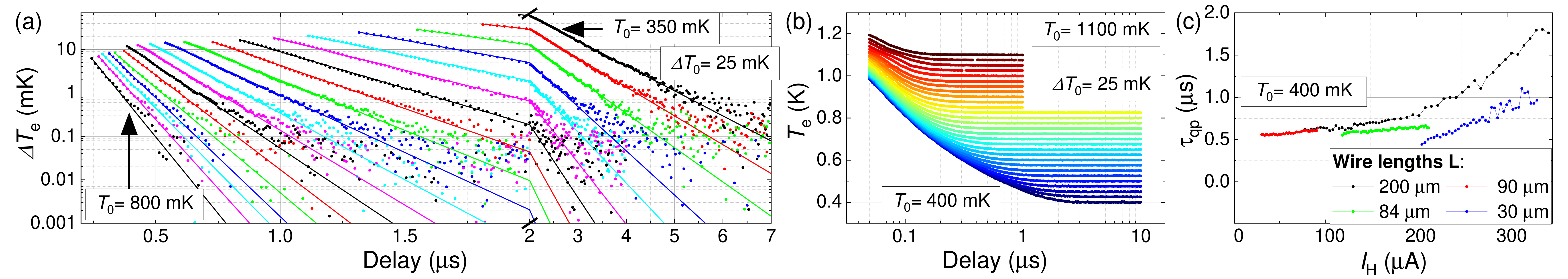}
\caption{Thermal relaxation of the aluminum nanowire. (a) Relaxations in the linear regime obtained from monitoring the temporal variation of the switching probability recorded for bath temperatures from $350$~mK to $800$~mK with step $\Delta T_{0}=25$~mK. The straight lines are fitted exponential decays. The horizontal scale changes at the $Delay=~2~\mu$s. (b) Relaxation in the wide range of temperatures recalculated from the temporal transient in the switching current recorded for bath temperatures from $400$~mK to $1100$~mK with step $\Delta T_{0}=25$~mK.
(c) Relaxation time versus heating current amplitude $I_H$ \cite{Supplemental} for constant bath temperature $T_0=400$~mK and various lengths $L$ of the wires.}
\label{Thermal_relaxation}
\end{figure*}

Our samples are superconducting nanowires fabricated from polycrystalline aluminum with elastic mean free path $l_{m}=15-20~$nm. Their thickness is $30~$nm, width - $600~$nm and lengths range from $L=37~\mu$m to $800~\mu$m. They are prepared by the standard single step electron beam litography on silicon substrates. In the middle of each nanowire there is a Dayem nanobridge (superconducting weak link), which serves as a fast and sensitive thermometer \cite{Supplemental}. Typically bridges are $150~$nm long and $80~$nm wide. In order to measure electron temperature of superconducting weak links in dynamical thermal transients, we use switching thermometry \cite{Zgirski2018, Zgirski2020} with the low-temperature operational limit of 280~mK \cite{Supplemental}. The temporal resolution is provided by a sequence consisting of two current pulses delayed with respect to each other \cite{Supplemental}: the heating pulse with amplitude $I_H$, intended to switch and overheat the structure above $T_c$, is followed by the testing pulse $I_t$ probing the temperature dependent switching probability of the nanobridge.

We analyze transients of temperature distribution in our samples using the 1-D heat transfer equation:

\begin{equation}
 C_e(T_e)\frac{\partial T_e}{\partial t}  = \frac{\partial }{\partial x}\left(\kappa(T_e)\frac{\partial T_e}{\partial x}\right)-P_{ep}(T_e,T_{ph})+P_{J} 
\label{eq:eq1}
\end{equation}
 where $T_e$ and $T_{ph}=T_0$ are electron and phonon temperatures respectively ($T_0$ is the bath temperature) and $C_e(T_e)$ is electronic heat capacity. The equation represents the conservation of energy: the change in the internal energy in a slice of the nanowire at a position $x$ is governed by the net heat flux due to diffusing QPs [the term with the electron thermal conductivity $\kappa(T_e)$], the heat flow from electrons to phonons $P_{ep}$ and - if a slice is at $T>T_c$ or biasing current exceeds the critical value - the Joule power $P_{J}$ generating new QPs. We take the experimentally determined temperature dependence of $C_e$ from literature \cite{Supplemental}, and numerically calculate $\kappa$ and $P_{ep}$ \cite{Supplemental}. Since $C_e$, $\kappa$ and $P_{ep}$ show strong temperature dependence the Eq.~\eqref{eq:eq1} can be solved only numerically. To initialize the calculation we impose a short current pulse (duration about 10 ns) with an amplitude exceeding the critical current of the bridge. It transits a large section of the wire into a normal state \cite{Supplemental}. We calculate the subsequent temporal evolution of the temperature profile of the nanowire. We get the relaxation time of QPs by fitting the exponential decay to the relaxation tail of the nanobridge temperature \cite{Supplemental}.

Considering Eq.~\eqref{eq:eq1} with no electric current running in the wire and neglecting the diffusion term, the thermal relaxation of QPs results only from electron-phonon coupling. In a regime of small electron temperature deviation from $T_0$, we approximate $P_{ep}\approx G_{ep}\Delta T_e$, where $\Delta T_e=T_e-T_{ph}$ is the difference between electron and phonon temperatures and $G_{ep}$ is the linearized thermal conductance between electrons and phonons, calculated numerically as a slope of heat flux $P_{ep}$ defined for small temperature differences near equilibrium, $G_{ep}=dP_{ep}/dT_e $ \cite{Supplemental}. The assumptions above lead to the electron temperature evolution described by a simplified equation:
\begin{equation}
 C_e(T_0)\frac{d \Delta T_e}{d t} = -G_{ep}(T_0)\Delta T_e
\label{eq:eq2}
\end{equation}
Both $G_{ep}$ and $C_e$  are assumed to depend only on the phonon (bath) temperature. The solution is thus an exponential decay with the electron-phonon relaxation time $\tau_{ep}(T_0)=C_e/G_{ep}$.
The electron-phonon coupling provides a theoretical upper bound for the expected relaxation time. The Eq.~\eqref{eq:eq2} is our reference cooling model, to which we further compare the experimental data and more detailed model [i.e. Eq.~\eqref{eq:eq1}] of the heat transfer.

The literature gives yet another prediction for the relaxation process resulting solely from the electron-phonon coupling \cite{Kaplan1976}:
\begin{equation}
 \frac{1 }{\tau_{ep}}=\frac{1 }{\tau_{0}}\sqrt{\pi} \left(\frac{2\Delta}{kT_c}\right)^{5/2}\sqrt{\frac{T}{T_c}}e^{-\frac{\Delta}{kT}}
\label{eq:Kaplan}
\end{equation}
where $\tau_0$ is material dependent electron phonon coupling time, $\Delta$ is the low temperature superconducting gap, $T_c$ is the critical temperature and $k$ is the Boltzmann constant. Eq.~\eqref{eq:Kaplan} provides an alternative determination of the QPs lifetime but it is valid only in the regime of low temperatures, where $T_{0}<T_c/10$. It was used in the previous experimental studies \cite{Wilson2001,Klapwijk2008,Klapwijk2011,deVisser2021}. It is confronted with our data and modeling in the experimental section of our paper.

\begin{figure*}[t]
\centering
\includegraphics[width=\textwidth]{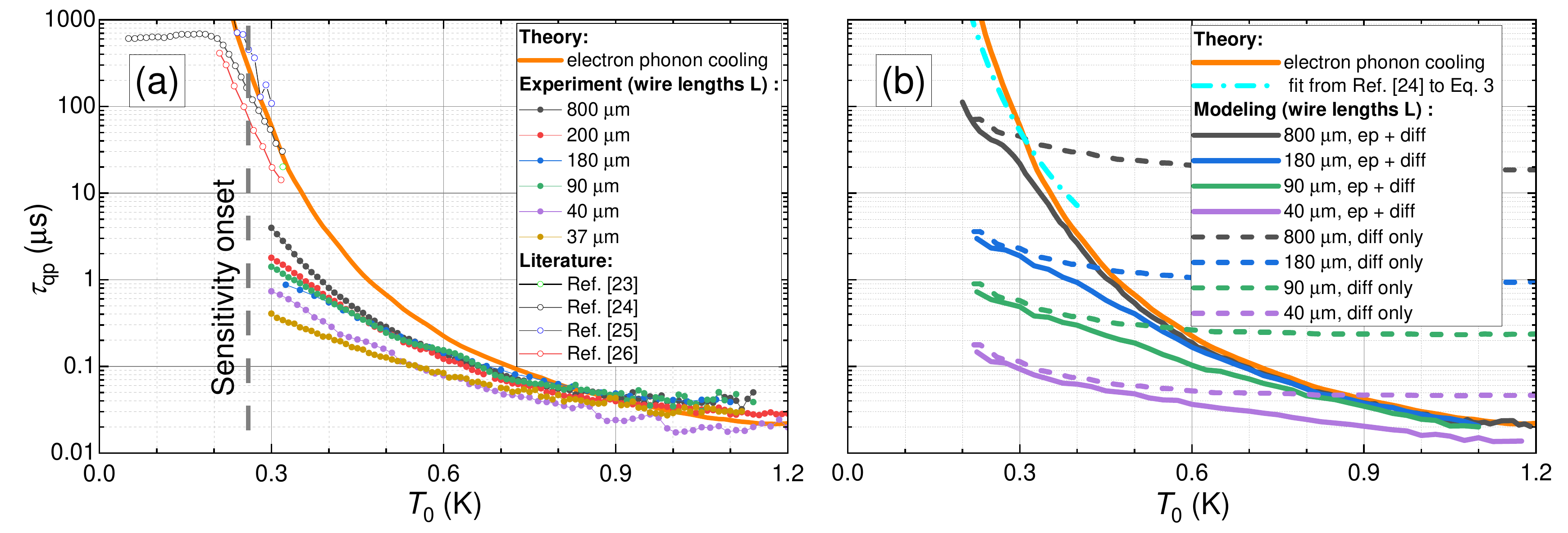}
\caption{Quasiparticle relaxation time $\tau_{qp}$ vs. bath temperature $T_{0}$. (a) Measurements (filled dots), literature data (open dots) and theoretical prediction resulting exclusively from electron-phonon coupling in the linear regime [solid orange line, Eq.~\eqref{eq:eq2}]. Curves displaying relaxation time below 10 $\mu$s for $T_{0}>0.3$~K are results measured by us for nanowires of various lengths, $L$. The data presenting $\tau_{qp}$ larger than $10~\mu s$ for $T_{0}<0.32$~K are extracted from literature. The dashed vertical line marks the operational limit of our thermometer \cite{Supplemental}. (b) The numerical solution of the heat flow equation [Eq.~\eqref{eq:eq1}] for 4 nanowires from the panel (a): the solutions taking into account both electron-phonon coupling and QPs diffusion are drawn with solid lines (label "ep + diff"). They branch out from the theoretical prediction resulting exclusively from electron-phonon coupling in the linear regime [solid, orange line, Eq.~\eqref{eq:eq2}, the same as in the left panel]. The longer the wire the lower temperature at which such separation occurs. According to modeling as the temperature is being lowered, the branching point marks the transition from cooling governed by the electron-phonon relaxation channel to that provided by pure QPs diffusion. In addition we provide low temperature prediction of the relaxation time reported in the literature [dashed-dotted line, Eq.~\eqref{eq:Kaplan}] and numerical solutions for cooling by diffusion only [Eq.~\eqref{eq:eq1} with $P_{ep}$=~0; dashed lines, label "diff only"].}
\label{tau_vs_length}
\end{figure*}

Fig.~\ref{Thermal_relaxation} shows the measurement of electron temperature relaxations in time domain for the wire of length $L=180 ~\mu$m recorded at various bath temperatures. The experiment is carried out in the two modes: by monitoring the temporal variation in the switching probability of the nanobridge at a fixed testing current [Fig.~\ref{Thermal_relaxation}(a)], and by measuring temporal profile of the switching current, defined as the one for which switching probability $P$=0.5 [Fig.~\ref{Thermal_relaxation}(b)]. The former mode delivers higher resolution, but it is only applicable for small departures from the bath temperature, while the latter allows to study the relaxation of QP in a wide range of temperature \cite{Zgirski2020}.

QPs lifetimes can be obtained from the experimental cooling curves by fitting the exponential tails of the relaxations. Both methods yield the same relaxation time $\tau_{qp}$ for a given bath temperature. Ideally, the QPs lifetime should not depend on the heating power. In Fig.~\ref{Thermal_relaxation}(c) we see that change in the heating current from $I_H=30 ~\mu$A to $I_H=200 ~\mu$A, i.e. more than 1~order of magnitude in the heating power, shows not essential effect on the measured $\tau_{qp}$. Hence we call the measured lifetimes the intrinsic ones. At higher heating powers, QPs lifetime starts to increase, presumably due to the high population of phonons trapped in the wire, resulting in $T_{ph}$ elevated above $T_0$. These phonons can be reabsorbed by Cooper pairs, and in consequence they can slow down the relaxation process \cite{RothwarfTaylor, Wilson2001}.

The relaxation curves, like that presented in Fig.~\ref{Thermal_relaxation}, collected for wires of various lengths allow to extract the corresponding QPs lifetimes $\tau_{qp}$ (Fig.~\ref{tau_vs_length}). It is the central result of our paper. We observe slow relaxations of the order of a microsecond at the lowest temperature of the experiment, i.e. $T_0=300$~mK. For high temperatures ($T_{0}>0.7$~K) we do not see any appreciable length dependence, suggesting that relaxation process is completely dominated by electron-phonon interaction. Here, the relaxation times flatten out and fall down to $\tau_{qp}=~60-20$~ns, still in reach for our switching thermometry, and remain in a remarkable agreement with those predicted by the theory, i.e. $\tau_{ep}=C_e/G_{ep}$ [cf.~Eq.~\eqref{eq:eq2}]. On the other hand, at lower temperatures all wires show the relaxation times significantly shorter than those expected from electron-phonon coupling only. This can, in principle, indicate the existence of a more efficient process responsible for the thermal relaxation. The natural candidate for such process is the QPs diffusion, which should be more pronounced in short wires. The span of experimentally measured relaxation times for various lengths at $T=0.3$~K is smaller than that given by the numerical solution of Eq.~\eqref{eq:eq1}. While in the experiment we get $\tau_{qp}=400$~ns for $L=37~\mu$m and for $L=800~\mu$m we have $\tau_{qp}=4~\mu$s, the corresponding relaxation times determined from heat flow equation are $\tau_{qp}=~100$~ns and $\tau_{qp}=23~\mu$s, respectively. The weaker length dependence observed in the experiment suggests that the electron thermal conductivity at low temperatures is suppressed more than the BCS theory predicts. For the short wires, whose relaxation times should be dominated by the diffusion process, we observe $\tau_{qp}$ longer than those predicted by ~Eq.~\eqref{eq:eq1} with $P_{ep}=0$ [Fig.~\ref{tau_vs_length}(b)]. For the long wire ($L=800~\mu m$), which should be hardly affected by the diffusion process, $\tau_{ep}(T)$ still runs significantly below the pure electron-phonon prediction [cf.~Eq.~\eqref{eq:eq2}]. 

This finding may signal a larger role of electron-phonon coupling and smaller efficiency of electron thermal conductivity at lower temperatures.  We show that by increasing theoretical values of $P_{ep}$ by factor of 8 and reducing $\kappa$ by factor of 4 we are able to find a quantitative agreement with the measured QPs lifetimes, simultaneously for short and long wires \cite{Supplemental}. However such good matching can be obtained only for a narrow range of bath temperatures. Alternatively, our measurement may mean that heat capacity is smaller than the value assumed in the modeling.

We compare our findings with the literature, which offers data at the temperatures $T_{0}~<~0.32$~K - we impose them on our experimental results in Fig.~\ref{tau_vs_length}(a). These data display relaxation times from 20 to $110~\mu$s at $T_{0}=0.3$~K, which is roughly 1 order of magnitude longer than what we find in our wires.
Partially, the difference can be ascertained to the elimination of the QPs diffusion in the referenced samples: Ref.~\cite{Klapwijk2008} measured QP relaxation in resonators galvanically decoupled from a feed line and Ref.~\cite{Wilson2001} used high gap leads. However, we notice that according to modeling the relaxation process in the middle of $L=800~\mu$m wire should be also hardly affected by diffusion. The other difference is the thickness of the samples. Ours are $30$~nm thick, while the quoted ones predominantly range from $100$~to~$200$~nm. The thicker films make it more difficult for phonons to escape into substrate after they were emitted during QPs recombination and the secondary QPs excitations due to phonon reabsorption are more likely \cite{RothwarfTaylor}. The escape process for phonons can be also affected by the type of used substrate and the quality of the interface between substrate and nanowire. They both affect the Kapitza resistance \cite{Semenov2006,Hakonen2017} and, as such, are responsible for overheating of $T_{ph}$ of nanowire over $T_{0}$. For a serious acoustic mismatch the evacuation of phonons from the film may be a bottleneck for the observed relaxation. Unlike the quoted references, our data point to the stronger electron-phonon interaction at lower temperatures fueling a faster recombination of QPs into Cooper pairs. This claim goes in line with suggestion found in Ref.~\cite{Pekola2009} which presents measurement of the electron temperature in a thermal steady-state and finds smaller overheating of electrons than implicated by the BCS-based recombination theory \cite{Supplemental}. Noteworthy, our calculation of pure electron-phonon cooling [Eq.~\eqref{eq:eq2}] agrees well with low-temperature prediction of Eq.~\eqref{eq:Kaplan} [Fig.~\ref{tau_vs_length}(b)], but it also covers the full temperature range of possible experiments.

\begin{figure}[]
\centering
\includegraphics[width=0.46\textwidth]{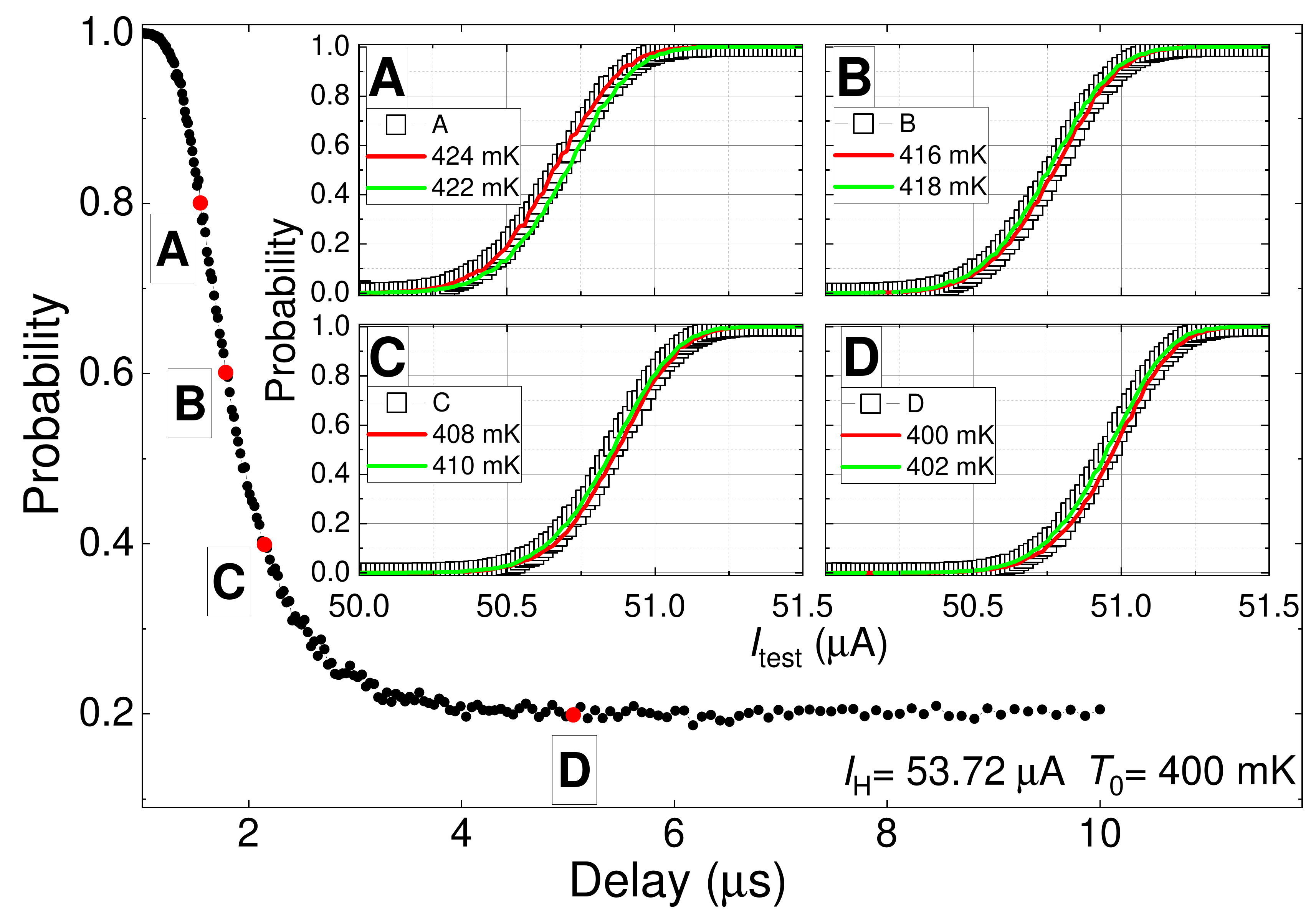}
\caption{The comparison of the switching probability dependencies (so called \textsf{S} curves) collected in the static and dynamic measurements. The main panel shows the temporal trace of the switching probability after exciting electrons with a short ($t_{H}=10$~ns, \cite{Supplemental}) heating pulse at $Delay=~0$. The insets show the dynamic \textsf{S} curves (open squares), collected for various delays corresponding to points A, B, C and D of the main trace, as assigned to the static \textsf{S} curves (solid lines) measured at the fixed bath temperatures displayed on the labels. If to assume that \textsf{S} curve is a unique fingerprint of quasiparticle occupation, the remarkable coincidence of the dynamic and static \textsf{S} curves implicates well-defined thermodynamic temperature of the electron gas overheated with respect to the lattice during thermal transient.}
\label{Static_vs_dynamic}
\end{figure}

One may wonder if the electron temperature $T_e$ is a properly defined concept during the rapid thermal transient considered in our experiments. Can we define the electron population with the Fermi-Dirac distribution which would be uniquely described by the parameter $T_e$ during such transients? Such apprehension makes some physicists use term of effective \cite{Pekola2016,Klapwijk2014,Clarke1994} or dynamic temperature \cite{Cleland2004PRB} to distinguish it from the thermodynamic temperature. For nonequilibrium states it is also common to use a microscopic approach which considers number and energies of quasiparticles explicitly and allows to avoid discussion of the electron temperature itself \cite{Nahum1998,Pannetier2009,Plourde2014,RothwarfTaylor}. The latter treatment, to ease the computation, very often involves many simplifications or assumptions, e.g. it neglects the energy dependence of the QPs density and the recombination rates. In equilibrium it may be fortunately replaced by thermodynamical approach, since in such case the temperature becomes the unique identifier of the number and energies of quasiparticles. In fact, the posed question is equivalent to asking what is the rate of electron-electron interaction $1/\tau_{ee}$ in a superconducting material. If it is much larger than the rate of electron-phonon relaxation $1/\tau_{ep}$, the temperature is properly defined. Electrons have enough time to "agree" the occupation of states themselves before they have chance to emit phonons. In the opposite limit, $\tau_{ep} \ll \tau_{ee}$, and electrons can be found in non-equilibrium occupation of states. The switching thermometry puts some experimental light on this problem. We measure exactly the same \textsf{S} curves in static and dynamic conditions (Fig.~\ref{Static_vs_dynamic}). This result suggests that the Joule-heated electrons converge very quickly to the Fermi-Dirac distribution, as expected for temperatures above $T_c=1.3~$K. When the heating current pulse is switched off, electrons start to lose energy cooling down towards the bath temperature $T_0=0.4~$K. It can be thought of as a quasi-static relaxation of the Fermi-Dirac distribution: electrons give up energy to phonons and the temperature defining the Fermi-Dirac distribution becomes lower as time progresses and QPs recombine back to the Cooper pairs. Thus, the electron temperature is a properly defined concept in the discussed experiments.

We directly measure QP lifetimes in superconducting aluminum in the temperature range $0.3~$K-$1.2~$K and present a complete thermal model to account for the data. The experimental relaxation times at $T_0>0.7~$K remain in quantitative agreement with the theory of electron-phonon coupling. A one order of magnitude difference in the theoretical and experimental QPs lifetimes at low temperatures ($T_0= 0.3~$K) suggests that the efficiency of the phonon emission is underrated in the theoretical model or there may exist an additional energy relaxation channel available for QPs. Our measurements point to the existence of quasi-equilibrium within an energy-relaxing QP system in a superconducting state, allowing to assign to the QPs a well-defined temperature during relaxation process.

\textit{Acknowledgments}\textemdash This work is financed by the Foundation for Polish Science project "Stochastic thermometry with Josephson junction down to nanosecond resolution" (First TEAM/2016-1/10) and National Science Centre Poland project "Thermodynamics of nanostructures at low temperatures" (Sonata Bis-9, No. 2019/34/E/ST3/00432). This research is partially supported by the "MagTop" project (FENG.02.01-IP.05-0028/23) carried out within the "International Research Agendas" programme of the Foundation for Polish Science co-financed by the European Union under the European Funds for Smart Economy 2021-2027 (FENG).

\end{document}